\documentclass[preprint,pteplogo]{ptephy_v2}

\preprintnumber{XXXX-XXXX} 
\usepackage{hyperref}

\usepackage{amsmath} 
\usepackage{amsthm} 
\usepackage{hyperref} 
\usepackage{graphics} 
\usepackage{algorithmic} 
\usepackage{url} 
\usepackage{amssymb}
\usepackage{graphicx}
\usepackage{siunitx}
\usepackage{geometry}
\usepackage{mathtools}

\begin{document}

\title{Modeling the Effects of Temperature and Electric Field on the Dark Count Rate of Photomultiplier Tubes}


\author{Yuto~Maekawa, Masamitsu~Fukazawa, Yasuhiro~Nishimura}
\affil{Graduate School of Science and Technology, Keio University \\
 3-14-1 Hiyoshi, Kohoku-ku, Yokohama, Kanagawa, Japan \email{yuto-mae@keio.jp}}

\begin{abstract}
Although photomultiplier tubes (PMTs) are sufficiently sensitive for applications based on counting single photons, the thermionic emission of electrons from the photocathode is one of the dominant background sources. In this study, we modeled the dark count rate of a given PMT as a function of the temperature and external electric field based on the Richardson–Dushman equation and incorporated the Schottky effect. To validate the model, measurements were performed using an eight-centimeter Hamamatsu R14374 PMT equipped with a bialkali photocathode. The dark count rate was measured over a range of temperatures from 10 to \SI{40}{\degreeCelsius}. An external electric field was applied to the outside surface of the optical window with the photocathode by varying the potential difference between the photocathode and the surrounding medium from 0 to 1100~V. The result confirmed that the influence of the external potential difference remained relatively small within a difference of 200~V, with the increase in the dark count rate remaining below 50~\% across all measured temperatures. This behavior was incorporated into the proposed model by including an error function to predict the dark count rate under operating conditions with varying temperatures and external electric fields.

\end{abstract}

\subjectindex{xxxx, xxx}

\maketitle

\section{Introduction}

Photomultiplier tubes~(PMTs) have been widely adopted in a variety of environments such as in combination with solid scintillators, water~\cite{Super-Kamiokande:2002weg,Hyper-Kamiokande:2018ofw}, ice~\cite{Aartsen_2017}, or liquid scintillators~\cite{juno2022juno,DUNE:2020lwj,ABE201378}. PMTs detect photons by converting them into photoelectrons at the photocathode before multiplying them through a chain of dynodes to produce a detectable signal.

The PMT performance is affected by various sources that contribute to the dark count noise, that is, to random false-positive detections of single photons. These contributions include thermionic emission of electrons from the photocathode, radiation-induced signals originating from radioactive impurities inside and outside the device, and time-correlated spurious pulses such as afterpulses~\cite{hamamatsuPMT}. Among these contributions, that of thermionic emission typically dominates the dark counts because thermoelectrons cannot be distinguished from photoelectrons. 

PMT performance is also influenced by a polarity of the operating high voltage ($\mathrm{HV}$). In the positive bias configuration ($\mathrm{+HV}$), the photocathode is grounded and the anode is biased at a positive high voltage. The photocathode remains at ground potential, which makes the noise picked up by the PMT less sensitive to variations in environmental factors such as temperature. However, a coupling capacitor is required at the output on such systems because the anode is at high voltage. As a result, the direct current readout is not appropriate and the circuit can distort the output waveform even with a pulsed input. In the negative configuration ($\mathrm{-HV}$), the anode is grounded while the photocathode is kept at a negative high voltage. Although this configuration enables direct signal readout at ground potential, it also induces a large potential difference between the photocathode and the surrounding medium. This potential energy generates an external electric field that enhances thermionic emission via the Schottky effect~\cite{schottky1926small}.

In this study, we investigated the performance of an 8-cm PMT (R14374, manufactured by Hamamatsu Photonics K.K.) developed with improved timing performance from the R12199 model (Hamamatsu). These PMTs were utilized in multi-PMT modules of the Hyper-Kamiokande~\cite{Hyper-Kamiokande:2018ofw} far detector and the intermediate water Cherenkov detector (IWCD)~\cite{scott2016intermediate}, as well as in the Cubic Kilometer Neutrino Telescope (KM3NeT)~\cite{adrian2016letter} with a $\mathrm{-HV}$ bias and in the outer detector of Hyper-Kamiokande with a $\mathrm{+HV}$ bias. Studies have shown that the dark count rate (dark rate) in case of $\mathrm{-HV}$ is more sensitive to changes in HV or temperature than with a positive voltage. Thus, even relatively slight fluctuations in the characteristics of the environment can make the rate unstable and affect the performance stability of low-energy detections or detectors with data acquisition using self triggering. Understanding how the performance of the PMT depends on the temperature and external electric field is essential to quantitatively characterize the response of detector systems and achieve a stable detection performance because the surrounding temperature can vary over time and with the position of the PMT.

Past studies have investigated how the dark rates of the PMT varied in terms of either temperature or electric fields separately~\cite{elorrieta2019characterisation}. To the best of our knowledge, no quantitative understanding of their combined impact in realistic operating conditions has been reported in the relevant literature.

To address this challenge, we propose a two-dimensional model of the dark rate that simultaneously incorporates both the temperature and the electric field. The model is based on the Richardson-Dushman equation~\cite{richardson1914lxvii,dushman1923electron} with a correction for the Schottky effect. It also includes an additional barrier correction described by an error function. The proposed framework provides a set of practical guidelines to predict the amount of noise generated by a PMT over a wide range of operating conditions. This estimation can then be applied to stabilize the overall performance of detector systems. 

\section{Theoretical Background}
The thermal electron current density of photocathode $J$ is described by the Richardson-Dushman equation as follows.
\begin{equation}
    J = A T^{2} \exp\left(\frac{-e\phi}{k_B T}\right),
\end{equation}
where $A=\frac{4\pi mek^{2}_{B}}{h^{3}} \approx 1.20\times 10^{6}~\mathrm{A~m^{-2}~K^{-2}}$ is the Richardson constant expressed in terms of the electron mass $m$ and Planck's constant $h$, $\phi$ is the work function of the photocathode material, $T$ is the absolute temperature, and $k_B$ is the Boltzmann constant.

The Schottky effect reduces the effective work function on the surface under an electric field $E$. The current density is modified as:
\begin{align}
J &= A T^{2} \exp\left(\frac{-e\left(\phi-\sqrt{\frac{e E}{4 \pi \varepsilon_0}}\right)}{k_B T}\right) 
= A T^{2} \exp\left(\frac{-e\phi_\mathrm{eff}}{k_B T}\right).
\label{eq:effphi}
\end{align}
Here, $\varepsilon_{0}$ is the permittivity of vacuum and $\phi_\mathrm{eff}$ is an effective work function. According to Eq.~\eqref{eq:effphi}, the temperature dependence of the dark rate $R$ is given as follows. 
\begin{equation}
    R=\alpha T^2\exp\left(\frac{-e\phi_\mathrm{eff}(E)}{k_BT}\right)+\mathrm{Const.},
\label{eq:darkcount}
\end{equation}
where $e$ is the elementary charge, $\alpha$ represents an effective emission scale incorporating the active emission area and electron collection efficiency, and the constant term ($\mathrm{Const.}$) accounts for the other non-thermal background components. Assuming that the dark rate follows this relation, it is expected to depend on both the temperature $T$ and the electric field $E$.

\section{Experimental Setup}
\begin{figure}[htbp]
        \centering
        \includegraphics[width=0.3\linewidth]{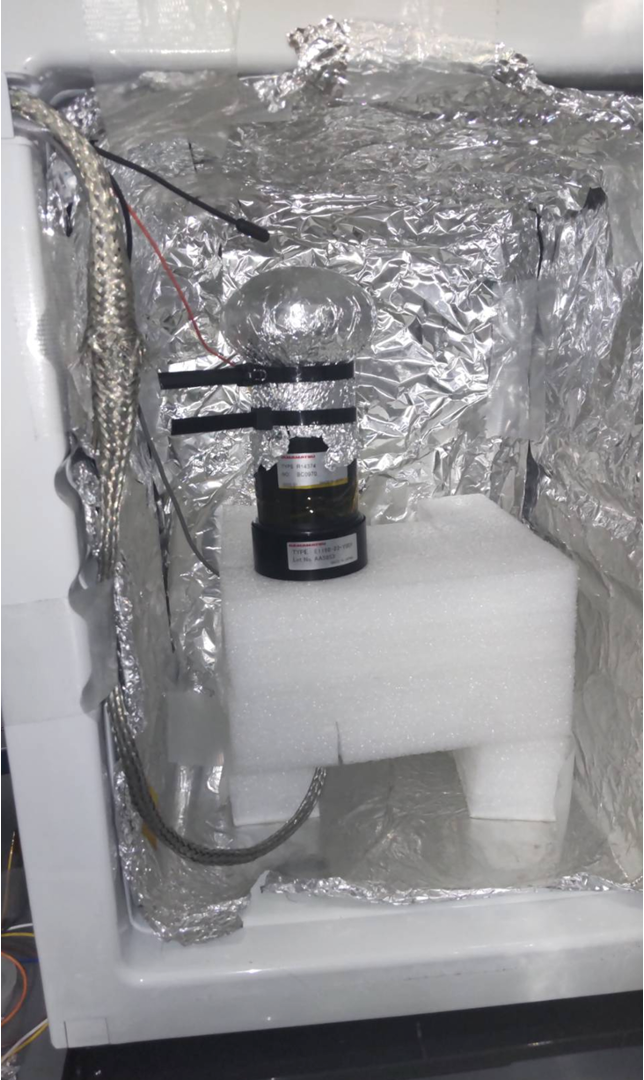}
        \includegraphics[width=0.6\linewidth]{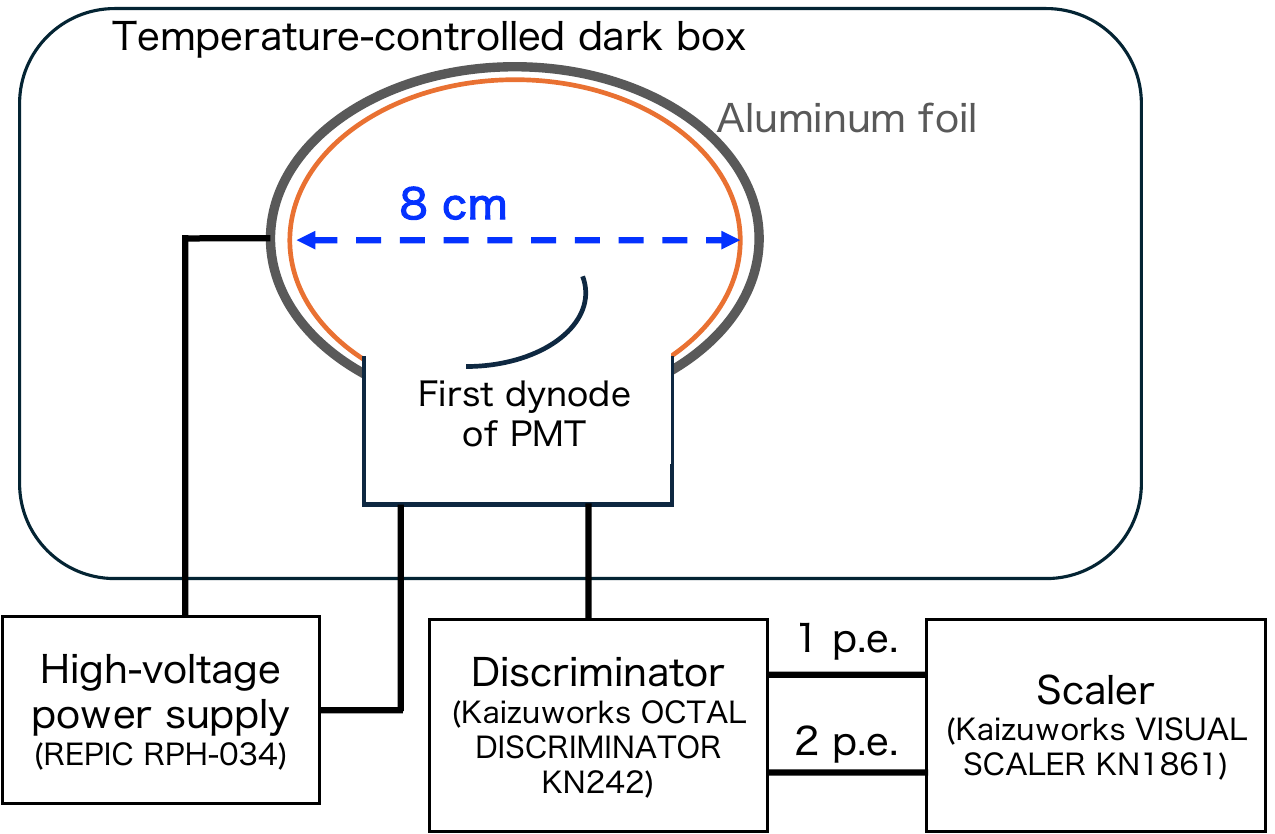} 
        \vspace{1mm}
    \caption{Left: A photograph of the measurement setup. Right: Schematic view of the circuit. A PMT with an 8-cm diameter was placed under a temperature control. An external voltage was applied to the aluminum foil covering the PMT.}
    \label{fig:setup}
\end{figure}

A temperature-controlled setup with adjustable external electric field was constructed as an ideal two-dimensional model of the dark rate. As shown in Fig.~\ref{fig:setup}, we evaluated an 8-cm R14374 PMT with a bialkali photocathode on a borosilicate glass window in a temperature-controlled dark box. A voltage of $-1100~\mathrm{V}$ was applied to the photocathode inside the box and the PMT was wrapped in aluminum foil with varying electric potential.

\begin{figure}[htbp]
        \centering
        \includegraphics[width=0.6\linewidth]{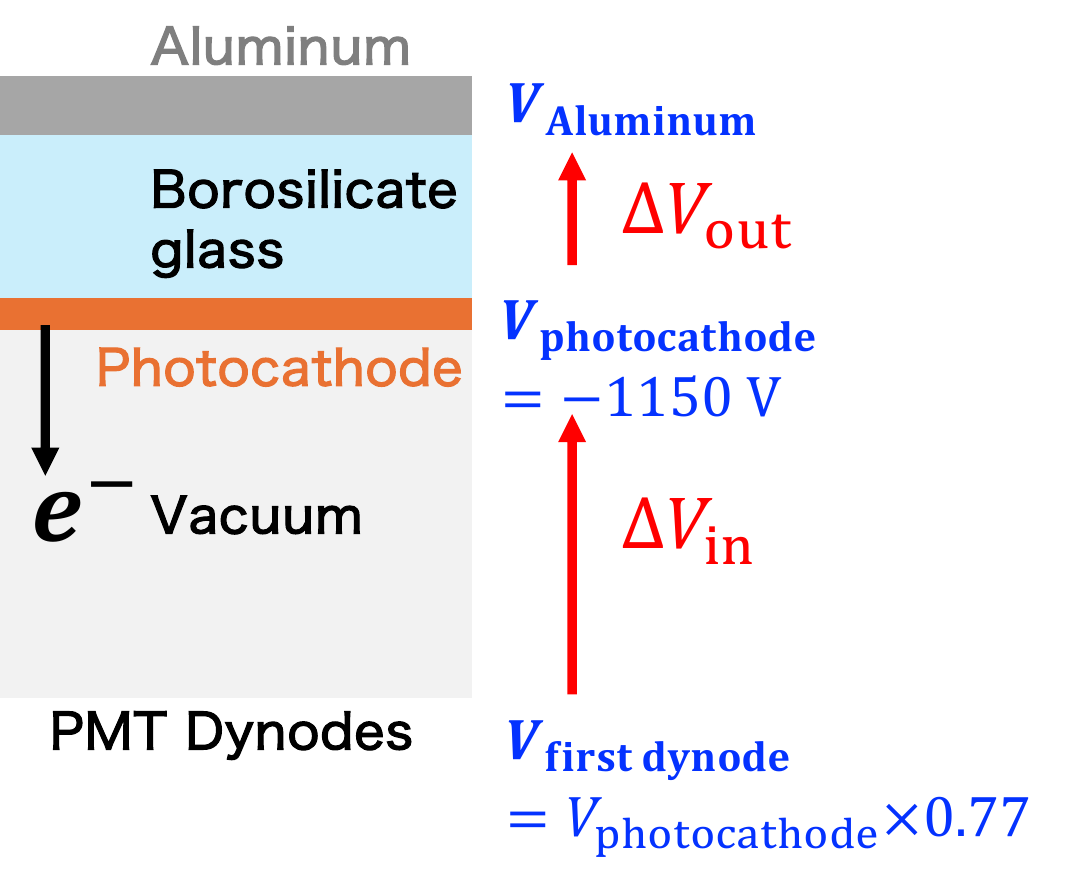}
        \vspace{1mm}
    \caption{Schematic view of electric potential at the PMT surface.}
    \label{fig:surface}
\end{figure}

Fig.~\ref{fig:surface} shows a schematic cross-section view of the electric potential at the surface of the PMT. The potential difference around the photocathode is represented by $\Delta V_\mathrm{out}$ and $\Delta V_\mathrm{in}$. A reference voltage $V_\mathrm{Aluminum}$ is applied to the aluminum foil covering the photocathode where the bias $V_\mathrm{photocathode}$ is supplied. The external potential difference $\Delta V_\mathrm{out}$ is obtained as $\Delta V_\mathrm{out} = V_\mathrm{Aluminum} - V_\mathrm{photocathode}$.

Electrons emitted from the photocathode are collected by the electric field between the photocathode and the first dynode of the PMT. The potential difference across this region is denoted as $\Delta V_\mathrm{in}$. The voltage applied to the first dynode $V_\mathrm{first~dynode}$ is proportional to the operating voltage of the PMT. This ratio was set to a value of $0.77$ determined by the PMT base circuit, which resulted in $\Delta V_\mathrm{in} = +253~\mathrm{V}$ under a PMT bias voltage of $-1100~\mathrm{V}$.

Measurements of the dark rate were carried out at a range of different temperatures from 10 to \SI{40}{\degreeCelsius} in increments of \SI{5}{\degreeCelsius} and for $\Delta V_{\mathrm{out}}$ values ranging from 0 to $+1100~\mathrm{V}$ in steps of $50~\mathrm{V}$ by varying $V_\mathrm{Aluminum}$. Meanwhile, the PMT dynode voltages and $\Delta V_\mathrm{in}$ are fixed during the measurement to maintain the same PMT detection performance except for the above environment. 

\begin{figure}[htbp]
    \centering
    \includegraphics[width=0.7\linewidth]{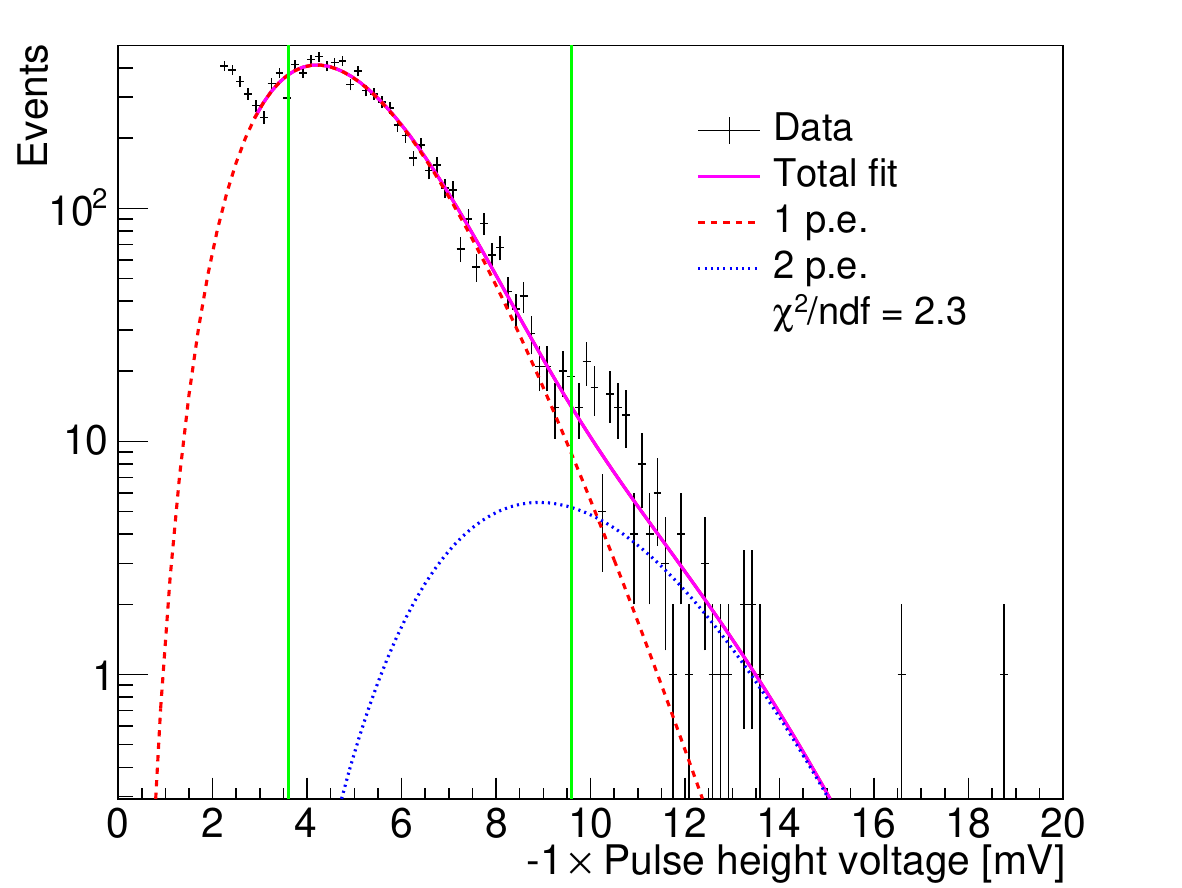}
    \caption{The pulse height distribution of the dark count signal with black points representing the measured data. The magenta solid line shows the total fit function, while the red dashed line and blue dotted line respectively correspond to the 1-p.e.\ and 2-p.e.\ contributions. The vertical light-green lines indicate the discriminator thresholds at $-3.6$ and $-$9.6~mV.}
    \label{fig:discri}
\end{figure}

Two discriminator thresholds were set with respective values of $-3.6$ and $-$9.6~mV for the single-(1-p.e.) and double-photoelectron (2-p.e.) levels to extract only thermal electrons at a 1-p.e.\ yield. The upper threshold was set to avoid counting undesired pulses over 1-p.e.\ by any light emission or external backgrounds. To estimate an efficiency for selecting 1-p.e.\ based on the hit counting provided by the PMT between the two pulse height thresholds, the pulse height distribution was recorded under the dark condition at 25~$^\circ$C as shown in Fig.~\ref{fig:discri}. The distribution was evaluated with the Polya function~\cite{PRESCOTT1966173} of $n$-p.e.\ signals $f_n(q)$, which is assumed to be
\begin{equation}
    f_n(q)=A_n\cdot\frac{a^a}{\Gamma(a)Qn}\left( \frac{q}{Qn}\right)^{a-1}\exp\left(-\frac{aq}{Qn}\right),
\end{equation}
where $A_n,~a$ are fitting parameters and $Q$ represents the mean pulse height of a 1-p.e.\ distribution. The results were fitted to the total distribution in a range from $-20$ to $-$2.8~mV using a combination of $f_1(q)$ and $f_2(q)$ to exclude 2-p.e.\ components. 

For the 1-p.e.\ component, the difference in the fractions of signals accepted by the two discriminator thresholds was estimated to be 74.5~\% from the fitted pulse-height distribution. This value was used as the 1-p.e.\ selection efficiency in the dark-rate calculation described in the following section. For the 2-p.e.\ component, 44~\% of the signals exceeded the 2-p.e.\ threshold, while the remaining 56~\% lay between the two thresholds. 

\section{Measuring the Dark Count Rate and Uncertainties}
The measured dark rate $R$ was calculated as follows.
\begin{equation}
    R=\left(N_\mathrm{1p.e.}-N_\mathrm{2p.e.}\right)/0.745,
    \label{eq:rate}
\end{equation}
where $N_\mathrm{1p.e.}$ and $N_\mathrm{2p.e.}$ represent event rates that respectively exceed the 1-p.e.\ ($-$3.6~mV) and 2-p.e.\ thresholds ($-$9.6~mV). Major uncertainties of instability and 2-p.e. contamination were evaluated as follows. 

\begin{figure}[b!]
    \centering
    \includegraphics[width=0.6\linewidth]{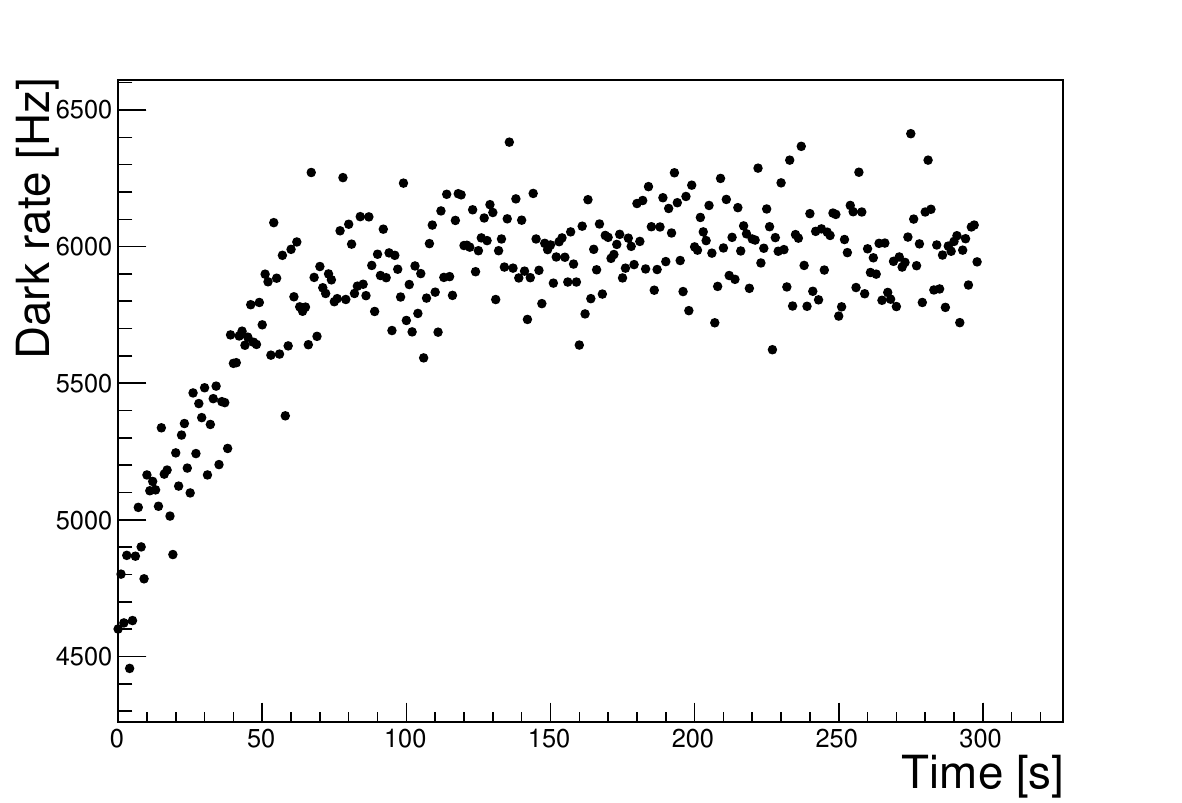}
    \caption{Time variation of a dark rate at $(T,~\Delta V_\mathrm{out})=(40~^\circ\mathrm{C},~550~\mathrm{V})$. The measurement started when the aluminum voltage setting was changed. Since the rate stabilized after about 100 s, the data in the remaining 200 s were used for the analysis.}
    \label{fig:timeplot}
\end{figure}

The dark counts were recorded every second for a total duration of 300 seconds after each setting change as shown in Fig.~\ref{fig:timeplot} for instance. An initial stabilization period of 100 seconds was excluded, and remaining set of 200 measurements was used for the analysis. At higher temperatures or larger $\Delta V_\mathrm{out}$, the dark rate tended to become unstable even after a relatively long period of time. Therefore, the standard deviation of these 200 measured points was taken as the uncertainty with respect to the dark rates to represent both statistical fluctuation and short-term instability. 

\begin{figure}[htbp]
    \centering
    \includegraphics[width=0.7\linewidth]{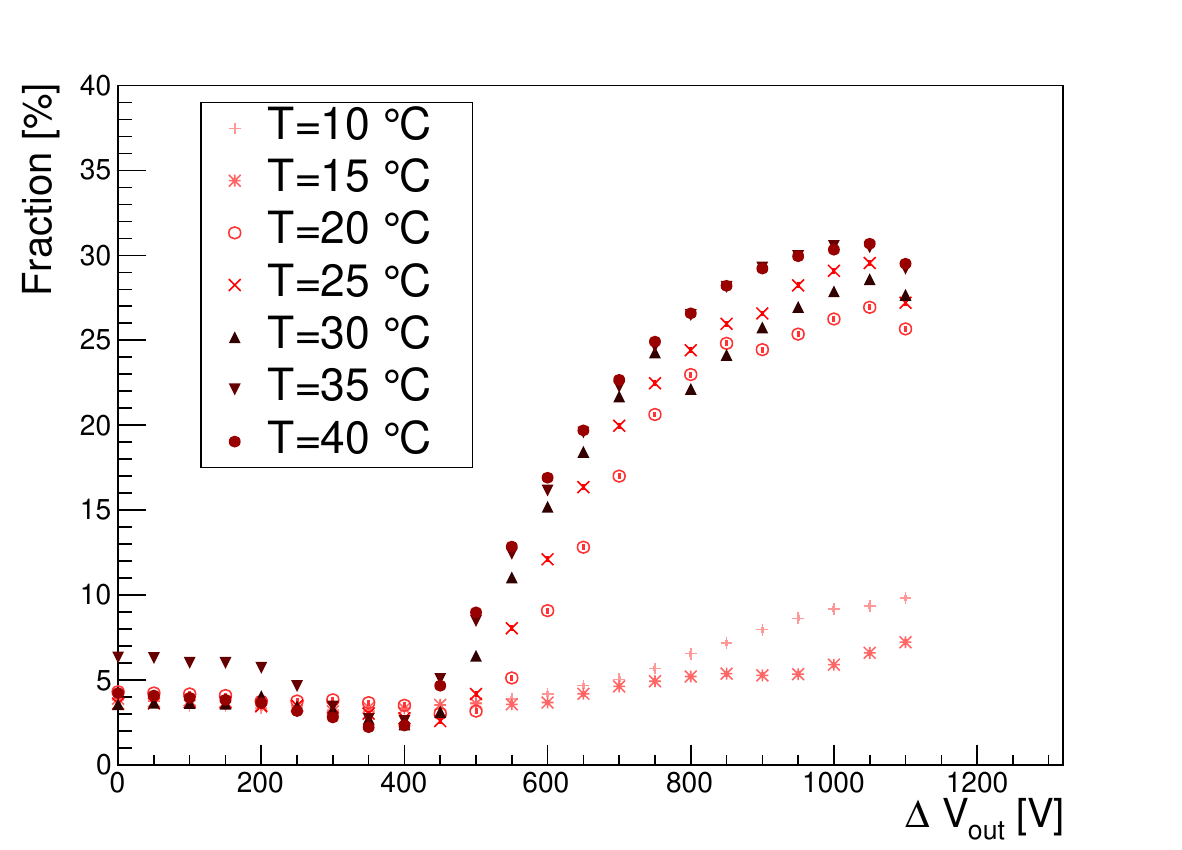}
    \caption{Dependence of the fraction of multi-photoelectron counts on $\Delta V_\mathrm{out}$ for each temperature setting. This fraction is defined as $N_{2\mathrm{p.e.}}/N_{1\mathrm{p.e.}}$.}
    \label{fig:fraction}
\end{figure}

The fraction of multi-photoelectron counts $N_{2\mathrm{p.e.}}/N_{1\mathrm{p.e.}}$ was investigated to confirm the stability of multi-photoelectron counting backgrounds. Fig.~\ref{fig:fraction} shows the dependence on $\Delta V_\mathrm{out}$ of the fraction for each temperature setting. The fraction remained relatively small in the low-$\Delta V_\mathrm{out}$ region, which indicates that most dark count signals consisted of single-photoelectron events. In contrast, the fraction increased significantly at larger $\Delta V_\mathrm{out}$, while remaining small at $10~^\circ\mathrm{C}$ and $15~^\circ\mathrm{C}$. Assuming the specified PMT pulse width of \SI{10}{ns} and a dark rate of 100~kHz, the probability of accidental pile-up was estimated to be 0.1~\%. This contribution of pile-up to the present measurement is therefore negligible.

The observed increase in the fraction suggests that contributions from signals of 2-p.e.\ or more largely emerged at high $\Delta V_\mathrm{out}$ at temperatures of $20~^\circ\mathrm{C}$ and higher. The 2-p.e.\ signals above the 2-p.e.\ threshold were canceled by the subtraction in Eq.~\eqref{eq:rate}. However, the 2-p.e.\ component lying between the two thresholds remained in the measured dark rate. This residual contribution can therefore cause an overestimation of the 1-p.e.\ dark rate. Although flashing by discharge around the aluminum foil or inside the PMT have been the source of this anomaly, this possibility has not been confirmed.

In the measurements at different $\Delta V_{\mathrm{out}}$ values, the uncertainty tended to increase at higher temperatures and larger $\Delta V_{\mathrm{out}}$ primarily because of the increased contamination of 2-p.e.\ signals. The lower and upper uncertainties for the dark rate results, $\sigma^{-}_R$ and $\sigma^{+}_R$, respectively, are therefore defined as
\begin{equation}
    \sigma_R^{-}
    =
    \sqrt{
        \sigma_{\mathrm{stat}}^2
        +
        \left(\Delta R_{\mathrm{2p.e.}}\right)^2
    },
    \qquad
    \sigma_R^{+}
    =
    \sigma_{\mathrm{stat}},
    \label{eq:total_uncertainty}
\end{equation}
where $\sigma_{\mathrm{stat}}$ and $\Delta R_{\mathrm{2p.e.}}$ are the statistical uncertainty and the contribution from the 2~p.e.\ contamination, respectively. From the fitted 2-p.e.\ component in Fig.~\ref{fig:discri}, 56~\% fell between the two discriminator thresholds, while the remaining 44~\% were above the 2-p.e.\ threshold. The possible contamination of the 2-p.e.\ component in the measured dark rate was conservatively estimated as $\Delta R_\mathrm{2p.e.}=0.56/(0.44 \times 0.745)~N_\mathrm{2p.e.}$.

\begin{figure}[htbp]
    \centering
        \includegraphics[width=0.7\linewidth]{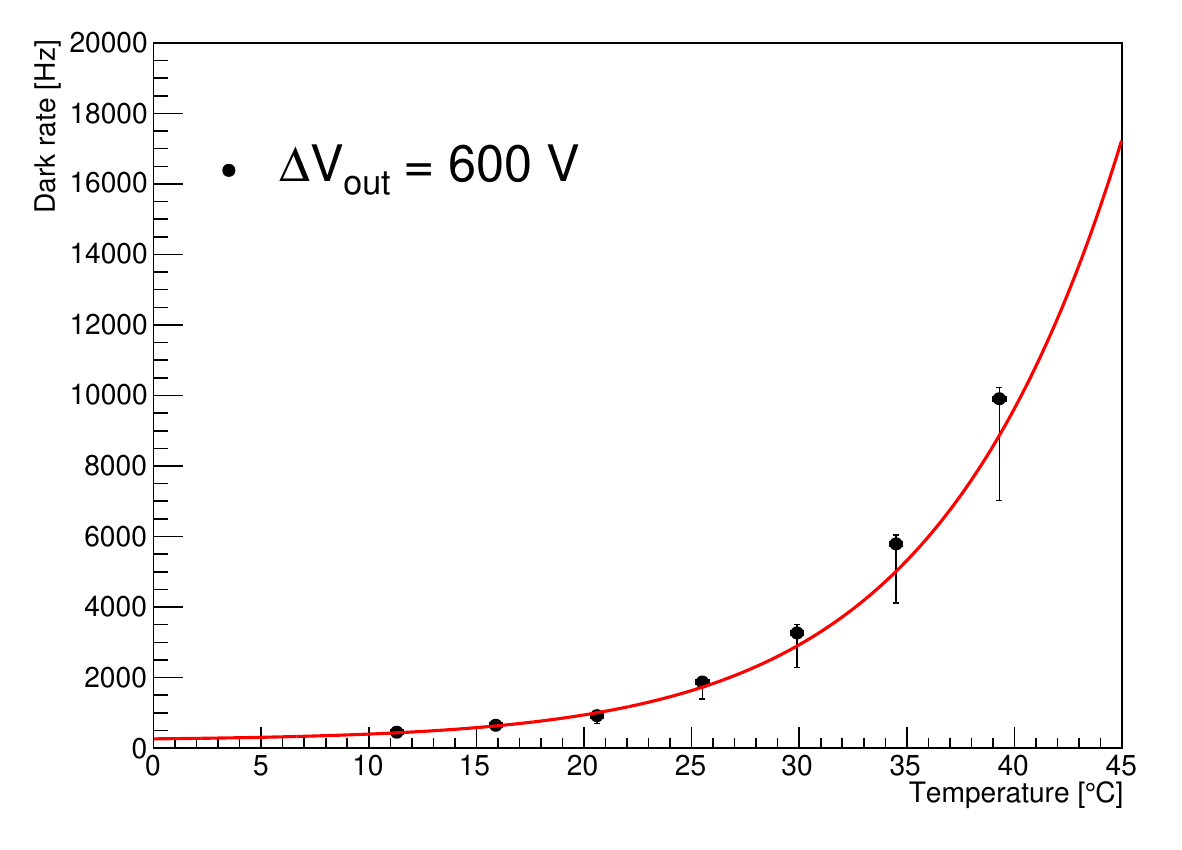}
        \label{fig:plot600V}
    \caption{The measured PMT dark rate data as a function of temperature at $\Delta V_\mathrm{out}=600~\mathrm{V}$. The black dots and the red line show the measured data and the fitted result obtained using Eq.~\eqref{eq:darkcount}.} 
    \label{fig:four_images}
\end{figure}

Fig.~\ref{fig:four_images} shows the measured dark rate as a function of temperature for a representative output voltage of 600~V. The dark rate exhibited a dependence on temperature at all voltages. 

\section{Result and Discussion}
The dependence of the dark rate on the temperature was quantified by fitting Eq.~\eqref{eq:darkcount} for each $\Delta V_{\mathrm{out}}$. The effective work function $\phi_\mathrm{eff}$ was extracted for each $\Delta V_\mathrm{out}$ setting based on the fitted data.
\begin{figure}[b!]
    \centering
    \includegraphics[width=0.6\linewidth]{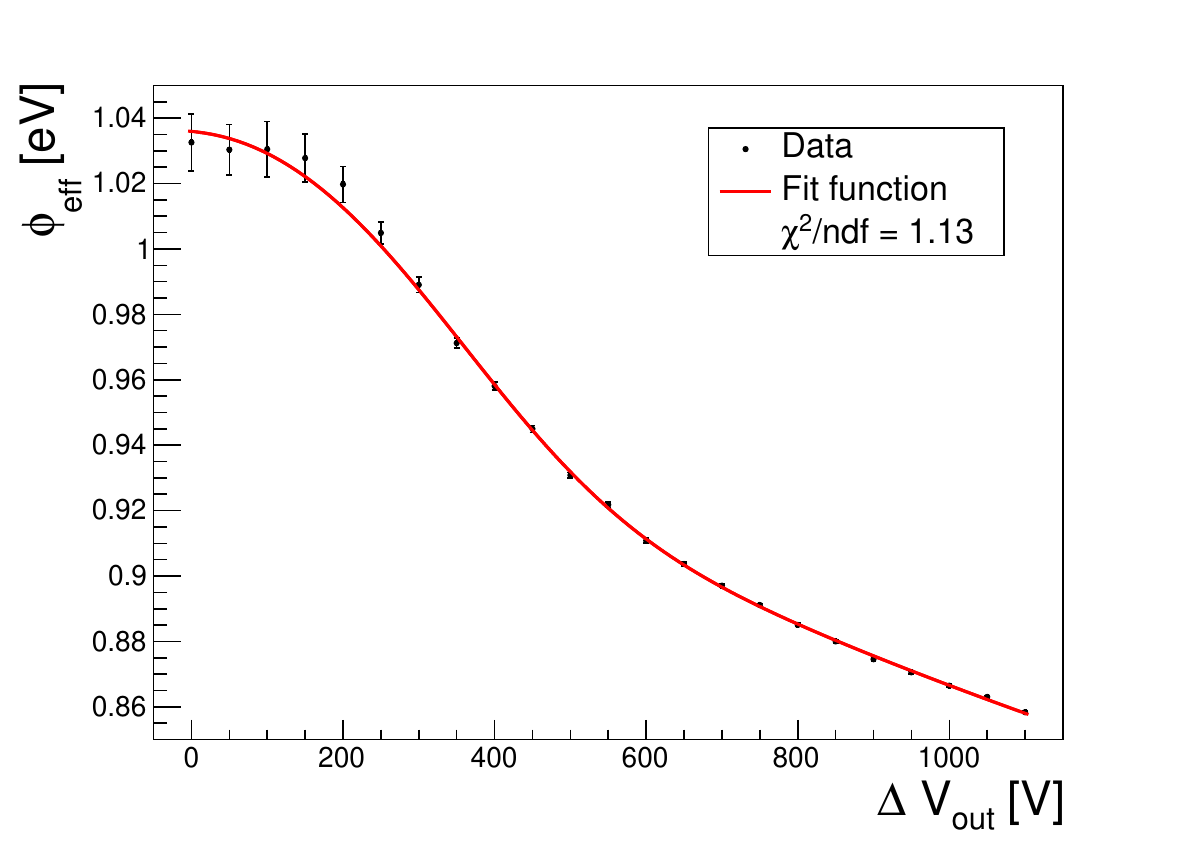}
    \caption{$\phi_\mathrm{eff}$ as a function of $\Delta V_\mathrm{out}$. The black points and the red line show the measured data and the fitted result using Eq.~\eqref{eq:efffit}, respectively.}
    \label{fig:effphi}
\end{figure}
Fig.~\ref{fig:effphi} shows the dependence of $\phi_\mathrm{eff}$ on $\Delta V_{\mathrm{out}}$. When a potential difference between the photocathode and the external environment was present, the electric field around the photocathode was composed of two contributions, including an external electric field through the optical window and the internal electric field in the vacuum. As expressed in Eq.~\eqref{eq:effphi}, due to the Schottky effect, the work function was effectively reduced by the electric field on photocathode. Using constant factors of $k_\mathrm{out}$ and $k_\mathrm{in}$ related to the PMT geometry, effective work function $\phi_\mathrm{eff}$ is parameterized as
\begin{equation}
\phi_\mathrm{eff}=\phi-\sqrt{k_\mathrm{out}\Delta V_\mathrm{out}+k_\mathrm{in}\Delta V_\mathrm{in}}.
\label{eq:phieff}
\end{equation}
From the shape of the figure, calculating these values confirmed that the effect of external-field on $\phi_\mathrm{eff}$ was small in the low-$\Delta V_\mathrm{out}$ region. This indicates that a modification of Eq.~\eqref{eq:phieff} was necessary to represent the data. As $\Delta V_\mathrm{out}$ was not able to reach the vacuum side due to a potential barrier of the thickness of the photocathode, an error function was introduced to Eq.~\eqref{eq:phieff} to make the function model more realistic. The expression for $\phi_\mathrm{eff}$ with the error function applied is thus given as
\begin{align}
    \phi_\mathrm{eff}&=\phi - \sqrt{k_\mathrm{out}' \Theta(\Delta V_\mathrm{out})\Delta V_\mathrm{out}+k_\mathrm{in}\Delta V_\mathrm{in}}~\label{eq:efffit},\\
    \Theta(\Delta V_\mathrm{out})&=\frac{1}{2}\left(1+\mathrm{erf}\left(\frac{\Delta V_\mathrm{out} - V_\mathrm{th}}{w_V}\right)\right)~\label{eq:efffit2}.
\end{align}
Here, $k_\mathrm{out}'$ is a constant parameter, and $\Theta(\Delta V_\mathrm{out})$ is a function composed of the error function. The parameter $V_\mathrm{th}$ represents the point at which $\Delta V_\mathrm{out}$ becomes effective at the vacuum side and $w_V$ is a parameter that characterizes the slope of the error function. The fit results obtained using the modified model given by Eq.~\eqref{eq:efffit} confirmed that the model reproduces the data successfully.

A single global fit to all dark-rate measurement points across all conditions was performed to determine the parameters in the modified $\phi_\mathrm{eff}$ model. A summary of the fit results is shown in Table~\ref{tab:finalresult}.
\begin{table}[!ht]
\caption{Fit results from a single global fit to all measurement points}
\label{tab:finalresult}
\centering
\resizebox{1.0\linewidth}{!}{
\begin{tabular}{ccccccc}
\hline
 $\alpha~[\mathrm{Hz}]$&$k_\mathrm{out}'[\mathrm{eV^2V^{-1}}]$ & $k_\mathrm{in}[\mathrm{eV^2V^{-1}}]$& $\phi~[\mathrm{eV}]$& $V_\mathrm{th}~[\mathrm{V}]$& $w_{V}~[\mathrm{V}]$&Constant [Hz]\\ 
\hline
 $(3.7\pm0.6)\times 10^{15}$&$(3.45\pm0.11)\times 10^{-5}$ & $(1.9\pm0.4)\times10^{-6}$ & $1.168\pm0.002$ & $349\pm13$&$244\pm14$&$258\pm6$\\
\hline
\end{tabular}
}
\end{table}

\begin{figure}[htbp]
    \centering
    \includegraphics[width=0.7\linewidth]{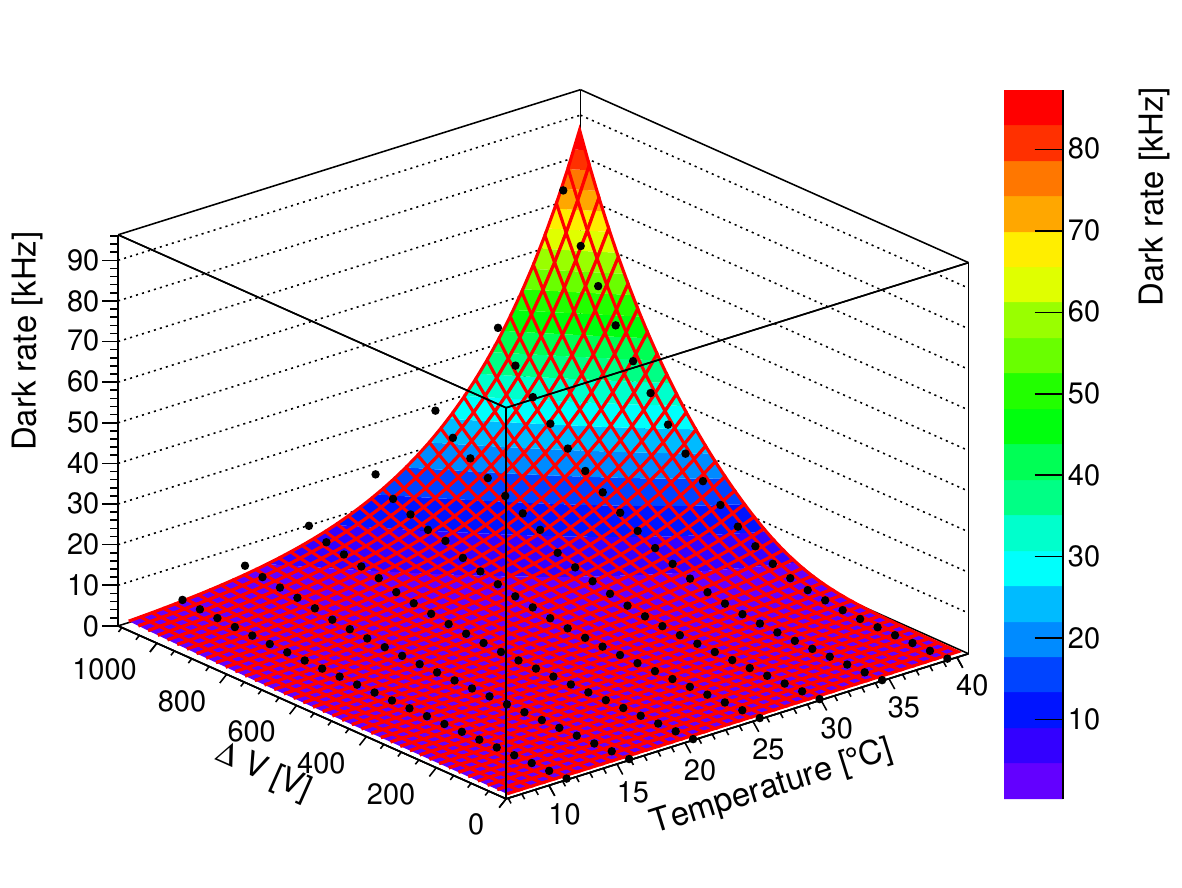}\\
    \includegraphics[width=0.49\linewidth]{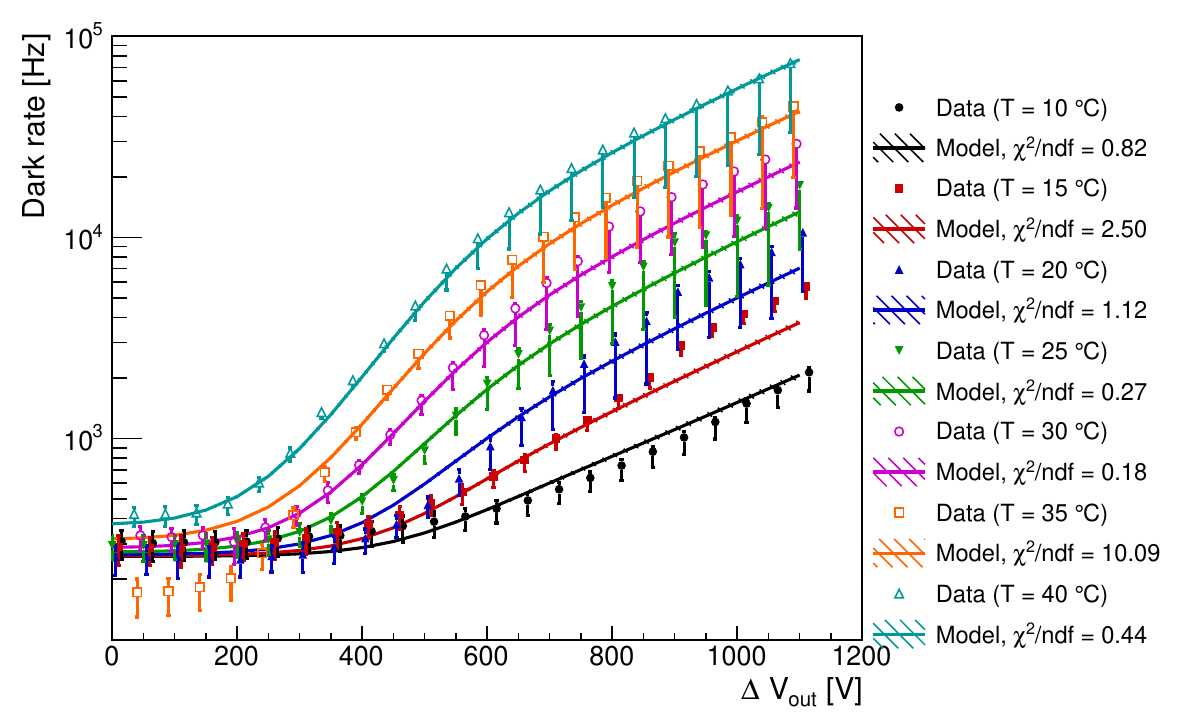}
    \includegraphics[width=0.49\linewidth]{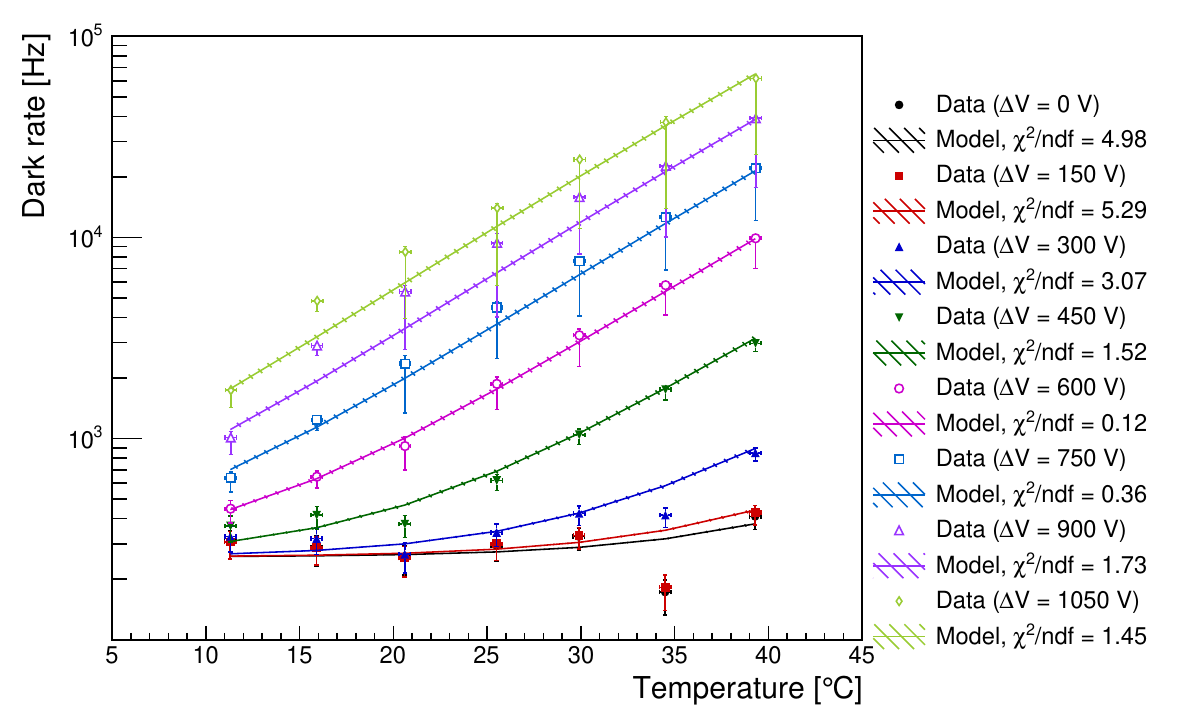}
    \caption{Top: Dark rate with respect to the $\Delta V_\mathrm{out}$ and temperature. Black dots show the measured data, and a colored mesh represents the model prediction by using the fitted parameters.
    Bottom left: Dark rate as a function of $\Delta V_\mathrm{out}$ for several temperatures. Colored dots show the measured data, and the lines represent the best-fit model. The hatched bands indicate the uncertainty of the model. 
    The $\chi^2/\mathrm{ndf}$ values for each temperature are calculated after the global fit. Data dots are slightly shifted along the horizontal axis for visibility. Bottom right: Dark rate as a function of temperature for several $\Delta V_\mathrm{out}$. Colored dots show the measured data, and the lines represent the best-fit model. The hatched bands indicate the uncertainty of the model.}
    \label{fig:slice}
\end{figure}

The top panel of Fig.~\ref{fig:slice} provides a three-dimensional representation to visualize the combined dependence on temperature and $\Delta V_\mathrm{out}$. The dependence of the dark rate on $\Delta V_\mathrm{out}$ and temperature is shown in Fig.~\ref{fig:slice}. The dark rate was modeled as a function of temperature and $\Delta V_\mathrm{out}$, and the measurements were described with a single unified expression. The observed discrepancies or large deviations in some measured points are attributed to instability arising from various sources such as fluctuations in temperature, variations in the external electric field, and burst-like noise in the photocathode. 

This fitted model indicates that the calculated rate at $\Delta V_\mathrm{out} \leq 200\mathrm{V}$ remained within a 50~\% increase relative to the value at $0~\mathrm{V}$ for all measured temperatures. The overall behavior over the two environmental parameters $\Delta V_\mathrm{out}$ and temperature was successfully represented by a single equation based on Eq.~\ref{eq:darkcount}, \ref{eq:efffit} and \ref{eq:efffit2}. 

\section{Conclusion}
The results demonstrate that the dark rate of the 8-cm PMT (Hamamatsu R14374) investigated in this study measured over $\Delta V_\mathrm{out}$ from 0 to +1100~V and temperatures ranging from 0 to 40~$^\circ$C were quantitatively described by the Richardson-Dushman model incorporating the Schottky effect. This provides a unified framework to understand the combined dependence on temperature and external potential difference, which had not been fully established in previous studies. The model allows a quantitative description of the dark rate as a function of temperature and external potential difference, and the findings suggests some strategies to mitigate dark noise effectively in future detector designs or to predict dark noise with a given environment. The small change in the dark rate in the low-$\Delta V_\mathrm{out}$ region indicates that reducing the potential difference between the photocathode and the surrounding environment in the model was effective in preventing a rapid increase in the dark rate. Of note, instability in the dark rate at high temperatures or large ambient potential differences from the photocathode can sometimes reduce the predictive accuracy of the proposed model. In conclusion, the proposed model provides a new mathematical description of temperature-induced instability in the PMT dark rate under varying electric fields around the photocathode relative to its voltage.
\clearpage
\section*{Acknowledgments}
This work was supported by JST SPRING (Grant Number JJPMJSP2123) and JSPS KAKENHI (Grant Numbers JP17H02885, JP20H01912).


%

\vspace{0.2cm}
\noindent

\bibliographystyle{ptephy}
\bibliography{main}

\let\doi\relax


\end{document}